\documentclass{emulateapj}
\usepackage{float}
\usepackage{amsmath}
\usepackage{epsfig}
\usepackage{graphicx}
\usepackage[]{subfigure}
\usepackage{color}
\definecolor{Blue}{rgb}{0.1,0.1,1.0}
\definecolor{Magenta}{rgb}{1.0,0.1,0.5}
\definecolor{LRed}{rgb}{0.8,0.0,0.0}

\newcommand{\nc}{\newcommand}
\nc{\cA}{{\cal A}}
\nc{\cR}{{\cal R}}
\nc{\bee}{\begin{equation}}
\nc{\ene}{\end{equation}}

\nc{\jcap}{JCAP}

\begin{document}

\title{Imprints of a hemispherical power asymmetry
  in the seven-year WMAP data due to non-commutativity of space-time}
\author{N.E. Groeneboom$^{1}$, M. Axelsson$^{1}$, D. F. Mota$^{1}$ and T. Koivisto$^{3}$}

\affil{$^{1}$ Institute of Theoretical Astrophysics, University of
  Oslo, P.O.\ Box 1029 Blindern, N-0315 Oslo, Norway\\
%$^{2}$ Centre of Mathematics for Applications, University of
%  Oslo, P.O.\ Box 1053 Blindern, N-0316 Oslo, Norway\\
$^{3}$ Institute for Theoretical Physics and Spinoza Institute, Utrecht University, 3508 Utrecht, Netherlands}

%\altaffiltext{4}{Texas Cosmology Center, The University of Texas at Austin, TX 78712}

\date{\today}

\begin{abstract}
  Non-commutative geometry at inflation can give arise to parity
  violating modulations of the primordial power spectrum.  We develop
  the statistical tools needed for investigating whether these
  modulations are evident in the Cosmic Microwave Background (CMB).
  The free parameters of the models are two directional parameters
  $(\theta,\phi)$, the signal amplitude $\cA_*$, and a tilt parameter
  $n_*$ that modulates correlation power on different scales. The
  signature of the model corresponds to a kind of hemispherical power
  asymmetry. When analyzing the 7-year WMAP data we find a weak
  signature for a preferred direction in the Q-, V-, and W bands with
  direction $(l,b) = (-225^\circ,-25^\circ) \pm (20^\circ, 20^\circ)$,
  which is close to another previously discovered hemispherical power
  asymmetry.  Although these results are intriguing, the significance
  of the detection in the W-, V- and Q-bands are nonzero at about $2
  \sigma$, suggesting that the simplest parameterization of the
  leading correction represents only partially the effects of the
  space-time non-commutativity possibly responsible for the
  hemispherical asymmetry.  Our constraints on the presence of a
  dipole are independent of its physical origin and prefer a
  blue-tilted spectral index $n_* \approx 0$ with the amplitude $\cA_*
  \approx 0.18$.

\end{abstract}


\maketitle

\section{Introduction}
\label{sec:introduction}

During recent years, studies of the cosmic microwave background (CMB)
have greatly improved our understanding of the early
universe. Observations of the CMB anisotropies, such as those obtained
by the Wilkinson Microwave Anisotropy Probe (WMAP) experiment
\citep{bennett:2003, hinshaw:2007}, have provided us with a deep
insight on the composition of structure and energy in our universe,
giving rise to the $\Lambda$CDM model. The $\Lambda$CDM model requires
that the universe has undergone an epoch of rapid accelerated
expansion. This epoch is named inflation, and is thought to have been
driven by a single or several scalar fields\citep{guth:1981}. In
addition, the model of inflation establishes a highly successful
theory for the formation of primordial density perturbations,
providing the required seeds for the large-scale structures
(LSS). These large scales structures are observed in the CMB today
\citep{guth:1981,linde:1982, muhkanov:1981, starobinsky:1982,
  linde:1983, linde:1994, smoot:1992, ruhl:2003, runyan:2003,
  scott:2003}.

Inflation can explain why the observed universe should be nearly
isotropic on large scales. However, anomalies found in the CMB during
the recent years \citep{de Oliveira-Costa:2004, vielva:2004,
  eriksen:2004a, groeneboom:2010, 2009ApJ...699..985H} suggest that
some anisotropy could be present at inflation too.

Several theoretical possibilities were put forward to explain such
anomalies \citep{ackerman:2007,ex1,ex2,exp3,exp4,ex5}. One of the
possibilities is to introduce a vector field that breaks the
rotational invariance in the early universe.  The presence of such
vector fields lead to quadrupole modulations of the CMB anisotropy,
which for vector perturbations was shown by \citet{durrer:1998}.
Recently, it was also suggested by \citet{groeneboom:2010} that the
5-year WMAP data contains a significant signal of a primordial vector
field which would break rotational invariance, corresponding to a
9$\sigma$ detection in the W-band . However, several authors have
since then claimed that the signal is due to systematic effects, as
the rotational axis is aligned with the rotational axis of the
satellite. A likely candidate for the source of the signal is
therefore asymmetric instrumental beams \citep{2010PhRvD..81j3003H,
  2010arXiv1001.4538K,bennet}. In addition, \cite{2010JCAP...05..027P}
have shown that there is no signal of breaking of
rotational-invariance in the observed LSS data, which makes less
likely that the signal is of a cosmological origin.

Unfortunately, such scenarios seem to be plagued with ghosts
\citep{Himmetoglu:2008hx} and several other sorts of instabilities
\citep{ins1,inst2,inst3}, but various generalizations and alternatives
can also be considered \cite{alt1,alt2,alt3,alt4}.

Several of the anomalies, such as the axis of evil or the
hemispherical asymmetry in the power spectrum, raise the question
whether there could exist odd-parity modulations in the CMB. It was
shown by \cite{2010arXiv1011.2126K} that the effects of space-time
non-commutativity at inflation can generate modulations. The leading
contribution, as generically in any parity-violating model, is a
dipole modulation, but up to now it hasn't been explicitly looked for
from the observed microwave sky.  There is a general framework for
describing the effects on the CMB anisotropy in terms of bipolar
harmonic expansion \citep{Hajian:2003qq,pullen:2007} which we we apply
here.
%In this paper, we consider this anisotropic model based on non-commutative theories. To the lowest order, the models give rise to
%a $\ell, \ell \pm 1$ dipole coupling in the covariance matrix, similar to the $\ell,\ell \pm 2$ quadrupole coupling as observed in the ACW signal.
The theoretical background is discussed in section
\ref{sec:background}, while the numerical implementation is discussed
in section \ref{sec:methods}. In section \ref{sec:properties}, we
investigate the model properties and conclude that the effect is
similar to a hemispherical dipole asymmetry. We perform several
analyses on simulated data and discuss the joint posterior results. In
section \ref{sec:analysis}, we analyze the seven-year WMAP sky maps
and find weak evidence for a preferred direction located at the
previously described hemispherical power asymmetry by
\cite{2009ApJ...699..985H}. In \ref{sec:conclusion} we conclude our
results, and discuss what how more general models might give rise to more
significant results.

\section{Background}
\label{sec:background}

The parity violating effect on the primordial power spectrum studied
in the present paper can be generated by quantum effects
during or shortly after inflation. Such effects have yet to be
considered due to the popular assumption that the power spectrum
should be invariant under spatial inversion. In principle there are no
physical arguments that exclude them as long as the physical
observables generated from the theory are consistent, so one should
therefore test this class of models against WMAP data and assess their
viability. The odd-parity bispectrum was recently explored in
a similar spirit by \cite{Kamionkowski:2010rb}.

It was shown in \cite{2010arXiv1011.2126K} that by taking into account
non-commutativity of space-time at very short distance scales, the
inflationary power spectrum becomes direction-dependent. This result
can be derived in several ways. The canonical $\theta$-type
non-commutation relations are \bee \label{com} \left[ \hat{x}_\mu,
  \hat{x}_\nu \right] = i\theta_{\mu\nu}\,, \ene where
$\theta_{\mu\nu}$ is an antisymmetric constant matrix of dimension
length squared that in the simplest case is a constant in a given
coordinate system. Effectively this results in a non-commutative field
theory. Another way to incorporate (stringy) uncertainty principle
into geometry is to define a deformation of the Heisenberg algebra of
the quantum fields. It was shown that with simple assumptions, in both
approaches one may predict, a power spectrum of the form\footnote{More
  generally, one obtains $P_{\theta}(\mathbf{k}) = P_{0}(k)[\alpha
  \cos({H \vec{\theta}\cdot \mathbf{k}}) + i\beta \sin({H
    \vec{\theta}\cdot \mathbf{k}})]$ where $\alpha$ and $\beta$ are
  parameters typically of order one, and the trigonometric functions
  can be replaced by hyperbolic ones depending on the
  prescription. However, at the leading order the behavior is
  qualitatively the same.}  \bee
\label{eq:pk}
P_{\theta}(\mathbf{k}) = P_{0}(k)e^{iH \vec{\theta}\cdot \mathbf{k}}
\ene
where $P_{0}(k)$ is the (isotropic) power spectrum when the non-commutativity is turned off. $H$ is
the Hubble factor and $\vec{\theta}$ is the time-space component of
the non-commutativity matrix,
%arising in the low-energy limit of open string theory
which sets the scale for the strength of the dipole modulation
correction, or $\cA_{1m} \sim |\vec{\theta}|$ in the present
formalism. For a comparison with other literature see the references
cited by \cite{2010arXiv1011.2126K}. The main feature of this model is
the parity-violating power spectrum referenced in eq. (\ref{eq:pk})
which implies that the two-point correlations, or the covariance
matrix, generates consistent physical observables such as CMB maps
without any further modifications.

However, it may turn out that similar criticisms apply also for these
early universe models as were mentioned against models which break rotation invariance at early times such as the one by \citet{ackerman:2007}. In particular,
the present realisations of field theories involving space-time non-commutativity may not be unitary. As a fully consistent
underlying non-commutative theory is still lacking, obviously we cannot make a unique and sharp prediction of the
fine details of its implications to cosmology. They depend upon the inflationary model and the form of non-commutativity (for example, whether $\theta^{\mu\nu}$ is constant in Eq.(\ref{com}) and in which coordinates).
These objections notwithstanding, our search of a dipole signature is strongly motivated since it is a generic prediction of these theories, and it seems a quite unique prediction too, since it appears to be otherwise difficult to realize the situation where the primordial spectrum of fluctuations is not symmetric. It is worth stressing also that order-of-magnitude estimates show that the theoretically plausible energy scales of non-commutativity are already probed by the present CMB experiments and the ensuing bounds will be pushed higher by the PLANCK satellite data. Thus, testing the leading order signature (dipole), gives us clues what to expect in a given non-commutative model for the next-to-leading signatures (higher multipole modulations and non-gaussianity). This provides an observational handle for the construction of viable and consistent theory.

%Albeit this is a pleasant feature when it comes to data analysis, the model itself is not without problems, again this is discussed by \cite{2010arXiv1011.2126K}.

The prediction for the CMB anisotropy pattern in non-commutative
geometry begins with the expansion of the primordial spectrum:
\begin{eqnarray}
\langle \cR(\mathbf{k}) \cR^{*}(\mathbf{k'}) \rangle & = & \nonumber
\delta^{3}(\mathbf{k}-\mathbf{k'}) \frac{2\pi^2 \sqrt{4\pi}}{k^3} \\ &\times&
\sum_{L,M}
\cA_{LM}\left(\frac{k}{k_0}\right)^{n_{LM}-1}Y_{LM}(\hat{k})
\end{eqnarray}
where the parameterisation by \cite{2009PhRvL.102c1301A} is
employed. The sum over $L$ could in principle cover the entire
spectrum of multipoles and $M$ runs from $-L$ to $L$. The resulting signal covariance is written as
\bee
\label{eq:scov}
S_{\ell m;\ell' m'} = \frac{i^{\ell-\ell'}}{2\pi^2}\sum_{L,M} \cA_{LM}
\, \xi^{LM}_{\ell m;\ell' m'} I^{LM}_{\ell \ell'} \ene where the
spectral-index $n_{LM}$ that parameterises the scale-dependence is
included in the integrated contribution over all scales, \bee
I^{LM}_{\ell \ell'} =
\int_{0}^{\infty}\frac{dk}{k}\left(\frac{k}{k_0}\right)^{n_{LM}-1}
\Theta_{\ell}(k) \Theta_{\ell'}(k)\,. \ene
The multipole moments of the sources $\Theta_{\ell}(k)$ are computed using a
modified version of CAMB \citep{lewis:2000} with the spectral index as
an input parameter. The geometrical factors $\xi_{\ell m; \ell' m'}$
in eq. (\ref{eq:scov}) are provided by \cite{2010arXiv1011.2126K}.  The
non-commutative nature of the fields responsible for the perturbations
give rise to non-hermitian signal covariance $S^{*}_{\ell' m'; \ell m}
= -S_{\ell m;\ell' m'}$ for the dipole ($L=1$) contribution which is
the only term in the expansion that is considered in the present
paper. The naive expectation from both theory and observations is that higher order terms are suppressed.
This make this assumption here but it should be examined more thoroughly.
The anisotropy in (\ref{eq:scov}) is
then added to the isotropic signal, corresponding to the $L=0$
term. One obtains a direct interpretation of $\cA_{00}\equiv \cA$ as
the primordial scalar amplitude in the canonical expression for the
isotropic matter spectrum. Furthermore \bee
\label{eq:apmz}
\vec \cA_{1M} = \cA_{1M} \hat r_M
%\cA_{1\pm 1} \propto \hat{r}_{\pm}, \quad \cA_{10} \propto \hat{r}_0
\ene where $A_{1M}$ is the amplitude which we are estimating. The
constant of proportionality in all three cases is $i/\sqrt{3}\,r
k_0 \cA$ where $r = |\mathbf{r}_0| = H\theta$. Here $\sqrt{\theta}$ is
the non-commutative lengthscale, $k_0$ is the pivot-scale.
The unit
vectors appearing in eq. (\ref{eq:apmz}) are the spherical vectors
parameterising the direction of the anisotropy, \bee
\label{eq:uvec}
\hat{r}_{\pm} = \mp \left(\frac{\hat{r}_x \mp
    i\hat{r}_y}{\sqrt{2}}\right), \quad \hat{r}_0 = \hat{r}_z\,.
\ene
In the case (\ref{com}) with $\theta$ constant in the comoving frame,
we have $\theta=|\vec{\theta}|$ and $n_{1M}=2$, but as argued above this need not be the case for general models.
Thus we will check this particular case separately but otherwise keep both the spectral index and the amplitude as free parameters.
So in all the model contains six unknown parameters, the three spectral
indices $n_{1M}$, the non-commutative lengthscale $\mu^{-1}$, and a
direction $(\theta,\phi)$ contained in the unit vectors of
eq. (\ref{eq:uvec}).  However, typically the slope of the modulation does not depend upon the azimuthal orientation. Then
the number of parameters is reduced to four: the direction
$(\theta,\phi)$, a coupling $\cA_*$ and the reduced spectral index $n_*$
for the $L=1$ case.

\section{Methods}
\label{sec:methods}
We now discuss the method for mapping out the desired posterior. The
method for obtaining the posterior is similar to the method presented in
\cite{groeneboom:2008b, groeneboom:2009a}, with the exception of a new
covariance matrix and a new parameter.  CMB data observations can be
modeled as:
\begin{equation}
  \mathbf d = \mathbf A \mathbf s + \mathbf n,
\end{equation}
where $\mathbf{d}$ represents the observed data, $\mathbf{A}$ denotes
convolution by an instrumental beam, $\textbf s(\theta,\phi) =
\sum_{\ell,m} a_{\ell m} Y_{\ell m}(\theta,\phi)$ is the CMB sky
signal represented in either harmonic or real space and $\textbf n$
is instrumental noise. It is generally a good approximation to assume both the CMB and noise
to be zero mean Gaussian distributed variates, with covariance
matrices $\mathbf{S}$ and $\mathbf{N}$, respectively. In harmonic
space, the signal covariance matrix is defined by $\textbf S_{\ell
  m,\ell'm'} = \left< a_{\ell m} a_{\ell 'm'}^* \right>$. In the
isotropic case, this matrix is diagonal. The connection to cosmological parameters
$\omega$ is made through this covariance matrix.  Finally, for
experiments such as WMAP, the noise is often assumed uncorrelated
between pixels, $\textbf N_{ij} = \sigma_{i}^2 \delta_{ij}$, for
pixels $i$ and $j$, and noise RMS equals to $\sigma_{i}$.

Let $\omega$ denote a set of cosmological parameters.
Our goal is to compute the full joint posterior $P(\omega |
\mathbf{d})$, which is given by $P(\omega
|\mathbf{d}) \propto P(\mathbf{d}| \omega ) P( \omega) = \mathcal{L}(
\omega) P(\omega),$ where $\mathcal{L}( \omega )$ is the likelihood
and $P(\omega)$ a prior. For a Gaussian data model, the likelihood
is expressed as:
\begin{equation}
  \mathcal L(\omega) \propto \frac{e^{-\frac{1}{2}\mathbf{d}^T
      \mathbf{C}^{-1}(\omega)\mathbf{d}}}{\sqrt{|\mathbf{C(\omega)}|}}.
\label{eq:likelihood}
\end{equation}
where $\mathbf{C}=\mathbf{S}+\mathbf{N}$ is the total covariance matrix.

\subsection{The Gibbs sampler}
The problem of extracting the cosmological signal $\mathbf s$ and
$\omega$ from the full signal by Gibbs sampling was addressed by
\citet{jewell:2004}, \citet{wandelt:2004} and
\citet{eriksen:2004b}. The CMB Gibbs sampler is an exact Monte Carlo
Markov chain (MCMC) method that assumes prior knowledge of the
conditional distributions in order to gain knowledge of the full joint
distribution. A significant fraction of the CMB data is completely
dominated by galactic foreground, and about $20\%$ of the data needs
to be removed. This might sound trivial, but in reality it complicates
processes as the spherical harmonics no longer are orthogonal. The
Gibbs sampler solves this problem intrinsically, as the galaxy mask
becomes a part of the framework \citep{groeneboom:2009b}.

The main motivation for introducing the CMB Gibbs sampler is
the drastically improvement in scaling. With conventional MCMC methods,
one needs to sample the angular power spectrum, $C_\ell= \langle
a_{\ell m} a_{\ell m}^* \rangle$, from the distribution  $P(C_\ell | \mathbf d)$,
which scales as $\mathcal O(N_{\textrm{pix}}^3)$, where $N_{\textrm{pix}}$ is the size of the covariance
matrix. For a white noise case, the Gibbs
sampler reduces this to $\mathcal O(N_{\textrm{pix}}^{1.5})$. In other
words, the Gibbs sampler enables effective sampling in the high-$\ell$ regime.

\subsection{Sampling scheme}

In order to sample from the full joint distribution
$P(C_\ell, \omega, \mathbf s |\mathbf d)$ using the Gibbs sampler, we
must know the exact conditional distributions $P(\mathbf s
| C_\ell, \omega, \mathbf  d)$ and $P(C_\ell, \omega | \mathbf s )$.
The Gibbs sampler then proceeds by alternating sampling from each of
these distributions:
\begin{align}
(C_\ell, \omega)^{i+1} \leftarrow& P(C_\ell, \omega| \mathbf{s}^i,
  \mathbf{d}) \\
\mathbf{s}^{i+1} \leftarrow& P(\mathbf{s} | (C_\ell, \omega)^{i+1}, \mathbf{d}).
\end{align}
The first conditional distribution is expressed as:
\begin{equation}
  P(C_\ell, \omega | \mathbf s, \mathbf d) = \frac{e^{-\frac{1}{2}\mathbf s^T\mathbf
      S(\omega)^{-1}\mathbf s}}{\sqrt{|\mathbf S(\omega)|}},
\label{eq:invgamma}
\end{equation}
and is distributed according to
an inverse Gamma function with $2\ell+1$ degrees of freedom. The remaining
conditional distribution is
\begin{equation}
  P(\mathbf s | C_\ell, \omega, d) \propto e^{-\frac{1}{2}(\mathbf s-\hat{\mathbf{s}})^T(\mathbf S(\omega)^{-1} + \mathbf N^{-1})(\mathbf s-\hat{\mathbf{s}})},
\end{equation}
where $\hat{\mathbf{s}} = \mathbf N^{-1}\mathbf d$. In other words,
$P(\mathbf s | C_\ell, \omega, \mathbf d)$ is a
Gaussian distribution with mean $\hat{\mathbf{s}}$ and covariance $(\mathbf S(\omega)^{-1} +
\mathbf N^{-1})^{-1}$. Numerical methods for sampling from these
distributions were discussed by \cite{groeneboom:2009b}.

\section{Model properties and predictions}
\label{sec:properties}
In this section, we review the numerical setup of the analysis. Most
of the framework is similar to the one employed by
\cite{groeneboom:2008b}. However, we introduce a new parameter
in addition to the anisotropic amplitude coupling $\cA_* \propto \cA_{00}$ and the
directional parameters $\theta$ and $\phi$. The parameter $n_*$
represents the spectral index of the correlation integral for the
$L=1$ case, and determines the tilt of the correlation power spectrum.

\subsection{Numerical setup}
\begin{figure}
\mbox{\epsfig{file=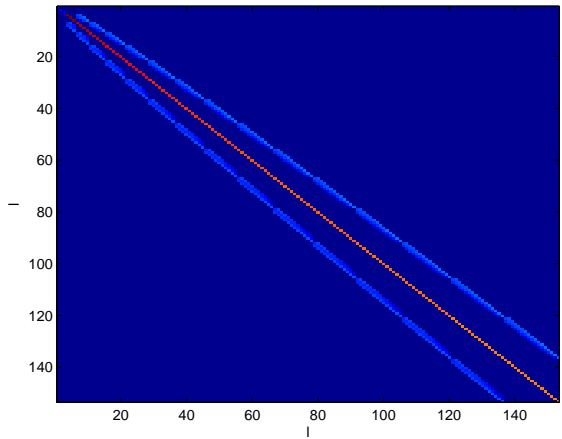, width=\linewidth,clip=}}
\caption{The non-commutative dipole-correlated covariance matrix
  $S$ on logarithmic scales. Blue represents lower values (or zero), red
  contains larger values.}
\label{fig:covmat}
\end{figure}
 We have previously developed a MCMC framework that enables
sampling over sparse anisotropic universe models, meaning models
predicting a covariance matrix that is sparse. This MCMC sampler is
integrated into a Gibbs sampler named COMMANDER
\citep{eriksen:2004b, eriksen:2008a}. The Gibbs sampler alternates
between sampling the CMB signal and the anisotropic model
parameters. The details on how this method was implemented was
described by \cite{groeneboom:2008b}.
\begin{figure}
\mbox{\epsfig{file=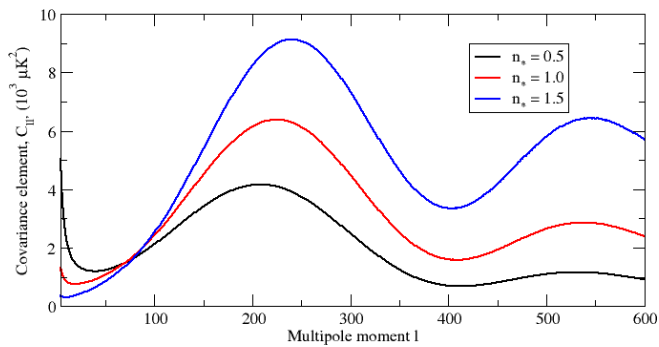, width=\linewidth,clip=}}
\caption{The transfer function Integrals $I_{\ell \ell'}^{1M}$ for
three values of $n_*$: $0.5$ (black), $1.0$ (red) and $1.5$
(blue). The integral correlates the power of the dipole signal for
various scales $\ell$, and are normalized according to their
area. Note that low values of $n_*$ corresponds to large-scale
correlations and vice versa, large values of $n_*$ corresponds to
small-scale dipole correlations.}

\label{fig:integrals}
\end{figure}

The covariance matrix is for this model given as follows:
\begin{equation}
  S= \langle a_{\ell m} a^*_{\ell' m'} \rangle  =
\frac{i^{l-l'}}{2\pi^2} \sum_{L=1}^{L=\infty} \sum_{M=-L}^{L}
\cA_{LM} \xi_{\ell m \ell' m'}^{LM} I_{\ell \ell'}^{LM}.
\end{equation}
Throughout this paper, we are only concerned with the dipole
modulations, or $L=1$. A plot of the
sparse covariance matrix $S$ can be seen in \ref{fig:covmat}.

The diagonal case $L=0$ represents the power spectrum $C_\ell$, where we
assume the tilt is equal to the spectral index $n_{lm} = n_s$ from the
best-fit seven-year WMAP spectral index \citep{2010arXiv1001.4635L}. We
introduce the parameter $n_*$ to replace $n_{lm}$ for the $L=1$ case as such:
\begin{equation}
I^{1}_{\ell \ell'} = \int_{0}^{\infty}\frac{dk}{k}\left(\frac{k}{k_0}\right)^{n_*-1} \Theta_{\ell}(k)
\Theta_{\ell'}(k)
\label{eq:integral}
\end{equation}

\begin{figure*}
\mbox{\epsfig{file=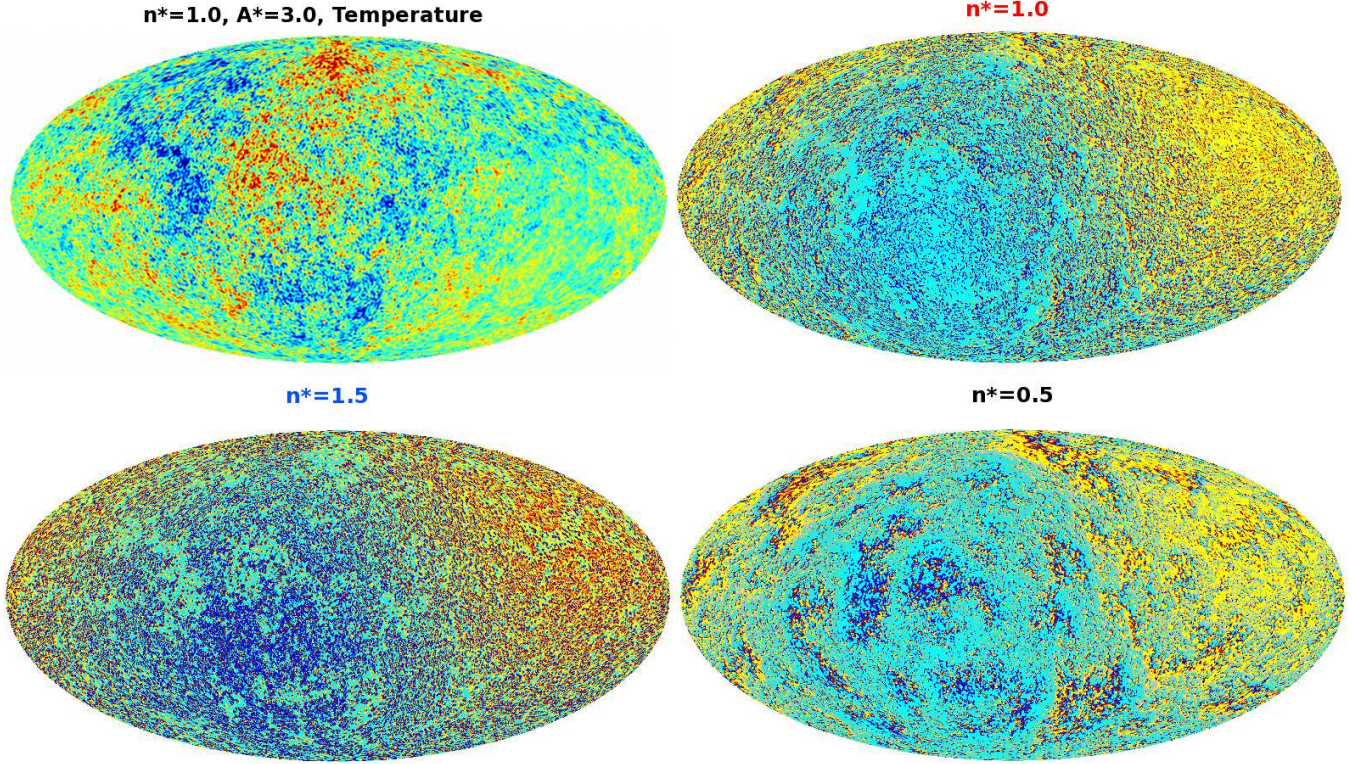, width=\linewidth,clip=}}
\caption{The effect of the dipole-modulating model on simulated CMB
  maps. Top left: A simulated CMB map with a large dipole
  contribution. The remaning maps show simulated dipole-modulated maps
divided by a non-dipole-modulated map. Note how different values
of the correlation tilt $n_*$ induce large-scale or small-scale dipole
correlations.}
\label{fig:dipolemap}
\end{figure*}

where we have chosen $k_0 = 0.05 \textrm{Mpc}^{-1}$ as the tilt scale.  Three
examples of $I^{1M}$ for $n_* =0.5, 1.0$ and $1.5$ are presented in
Figure \ref{fig:integrals}. The integral in equation \ref{eq:integral}
needs to be pre-computed in numerical software such as CAMB
\citep{lewis:2000}. It is difficult to implement CAMB into the
existing Gibbs sampler framework, so we utilize a different
scheme. First, we pre-compute 10 000 integrals for $n_*$ in an
interval $[-1, 4]$ and store the data in a binary file. This interval
is large enough to allow for all types of scale-dependence in the
anisotropy, and is our prior for $n_*$. For all purposes we consider
$n_*$ to be continuous. The precomputed binary file is then loaded
into the anisotropic MCMC framework such that $n_*$ can be treated as
a free parameter.

When simulating CMB maps for this model, the connection to a
hemispherical dipole asymmetry becomes imminent. This is depicted the
top-left frame of Figure \ref{fig:dipolemap}, where we present a
simulated CMB map with $\cA_* = 3.0$ and $n_*=1.0$ with direction $(l,b)
= (224^\circ, – 22^\circ)$. In order to verify that this really is a
signal similar to a dipole modulation, we simulate a map with
$\cA_*=0.0$ and divide the $A_*=1.0$ map with the $\cA_*=0$ map. The
resulting dipole structure can be seen in the remaining tree maps in
Figure \ref{fig:dipolemap}, where we have included simulations with tilt parameter $n_*=0.5$ and
$n_*=1.5$. Note that from the integral in Figure
\ref{fig:integrals}, $n_*=0.5$ corresponds to a large-scale dipole
modulation while $n_*=1.5$ contains more small-scale modulations.

\subsection{Analyzing simulated data}

\begin{figure}
\mbox{\epsfig{file=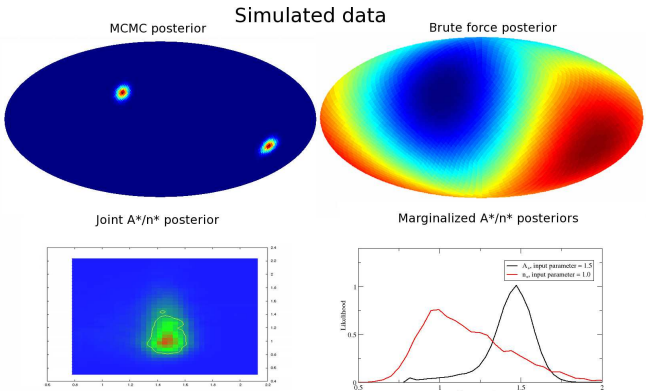, width=\linewidth,clip=}}
\caption{Results from a simple analysis of simulated data without
  foregrounds, noise or convolution. The input parameters
  were $\cA_* = 1.5, n_* = 1.0$ with direction $(l,b) = (224^\circ, –
  -22^\circ)$. Top left: The directional posterior from a MCMC
  sampling. Top right: the directional posterior from a brute-force
  run. Lower left: The joint posterior $P(\cA_*, n_* |
  \mathbf{d})$. Note that these parameters are not degenerate. Lower
  right: the marginal posteriors $P(\cA_*|\mathbf{d})$ and $P(n_* |
  \mathbf{d})$. }
\label{fig:simulated_results}
\end{figure}

\begin{figure}
\mbox{\epsfig{file=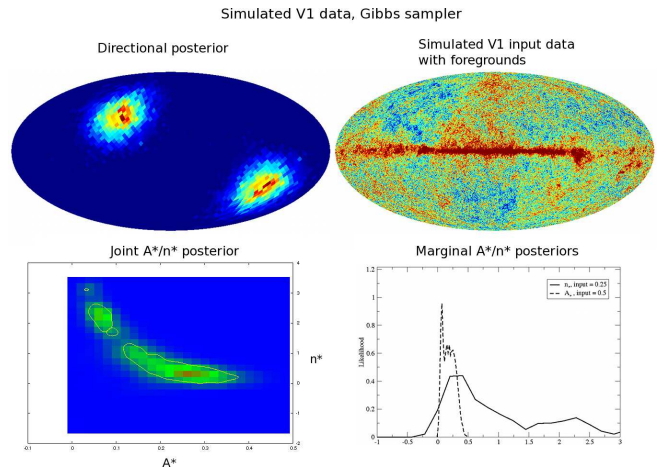, width=\linewidth,clip=}}
\caption{Results from a full-scale analysis of simulated realistic
  WMAP V-band data including foregrounds, noise and convolution. The
  input parameters were $\cA_* = 0.25, n_* = 0.25$ with direction $(l,b)
  = (224^\circ , -22^\circ)$. Top left: The directional posterior
  $P(\hat n | \mathbf {d})$.  Top right: The simulated V1 map with
  foregrounds and a dipole modulation.  Lower left: The joint
  posterior $P(\cA_*, n_* | \mathbf{d})$. Note that these parameters are
  more degenerate for low values of $\cA_*$. Lower right: the marginal posteriors
  $P(\cA_*|\mathbf{d})$ and $P(n_* | \mathbf{d})$. }
\label{fig:simulated_V1_results}
\end{figure}

We verify code by performing both a brute-force and MCMC analysis on a
noise-free, unconvolved simulated CMB map with no sky cut. In order to
build the joint two-dimensional distribution of $n_*$ and $\cA_*$, we
need an increased amount of samples as compared to the case by
\cite{groeneboom:2008b}. In addition, adding a new parameter will in
general decrease the significance of the results if the parameters are
correlated. As $\cA_*$ and $n_*$ can be expected to be correlated, we
choose to normalize the integrals such that the area under each graph
for all $n_*$ are the same. The degeneracy is broken, as can be seen
in Figure \ref{fig:simulated_results}. Here, we create a noiseless,
unconvolved CMB map with $n_{\textrm{side}}=128$ and $\ell_{\textrm{max}} =
\ell_{\mathrm{max}}^{\mathrm{cutoff}}=150$ adopting the best-fit $\Lambda$CDM model
determined from the seven-year WMAP data
\citep{2010arXiv1001.4635L}. The model input parameters are
$n_*=1.0$, $\cA_*=1.5$ and $(l,b) = (224^\circ, – 22^\circ)$. The joint
and marginal posteriors are presented in Figure
\ref{fig:simulated_results}. Note how the posterior of $n_*$ is
distributed similarly to a $\chi^2$-alike distribution, and that $\cA_*$
and $n_*$ are not degenerate. The distribution is always symmetric
around $\cA_*=0.0$, due to negative amplitude $\cA_*$ in a direction
$(l,b)$ corresponds to a positive amplitude in the opposite direction
$(-l, -b)$. This is not a problem when $\cA_*$ is large when compared to
its standard deviation, but for low values of $\cA_*$ it becomes
difficult to separate the negative peak from the positive, as they
will merge.

We continue by creating realistic V-band differencing assembly (DA) simulations with
$n_{\textrm{nside}}=512$, $\cA_*=0.25$, $n_*=0.25$ with direction
$(l,b)=(224^\circ, -22^\circ)$. The maps are produced using the V-band beam
and noise properties. In addition, we add synchrotron, free-free and
thermal dust foreground templates as described by \cite{2010arXiv1001.4555G}. The
V1 simulation is depicted in Figure \ref{fig:simulated_V1_results}. The
analysis is performed using the Gibbs sampler for
$l_{\textrm{max}}^{\textrm{cutoff}} = 400$ where we impose the KQ85 mask
\citep{gold:2008}, which removes 18\% of the sky. The analysis successfully
reproduce the input parameters, as is seen in Figure
\ref{fig:simulated_V1_results}. However, note that the distributions are wider
than in the noiseless, perfect case, and that tail of the marginal posterior
$P(A_* | \mathbf{d})$ merges with the positive values of $P(-A_* | \mathbf{d})$ near
zero.

\section{WMAP analysis}
\label{sec:analysis}
In this paper, we consider the seven-year WMAP temperature sky maps
\citep{2010arXiv1001.4744J}, and analyze the V-, W and Q (61, 94 and
41 GHZ, respectively). The V- and W-bands are believed to be the
cleanest WMAP bands in terms of residual foregrounds. We adopt the
template-corrected, foreground reduced maps recommended by the WMAP
team for cosmological analysis, and impose the KQ85 mask
\citep{gold:2008}. Point source cuts are imposed in all cases.  We
analyze the data frequency-by-frequency, and consider the combinations
V1+V2, Q1+Q2 and W1 through W4. The noise RMS patterns and beam
profiles are taken into account for each DA individually. The noise is
assumed uncorrelated. For details on joint Gibbs analysis of
multi-frequency data, see \citet{eriksen:2004b}. All data used in this
analysis are available from LAMBDA.

The angular resolutions of the V-, W- and Q bands are $0.35^{\circ}$,
$0.22^{\circ}$ and $0.53^\circ$, respectively. The sky maps are
pixelized at a HEALPix resolution of $N_{\textrm{side}} = 512$ with
$7'$ pixels. We adopt a harmonic space cutoff of
$\ell_{\textrm{max}} = 800$ for the two data sets, probing
partly into the noise dominated regime. However, we do not consider
multipoles at $\ell > 400$ for the anisotropic part of the signal
covariance matrix, in order to minimize the chance of systematic
effects such as residual point source contributions, beam
uncertainties or noise mis-estimation to affect our results.

\subsection{Results}
\begin{figure*}
\mbox{\epsfig{file=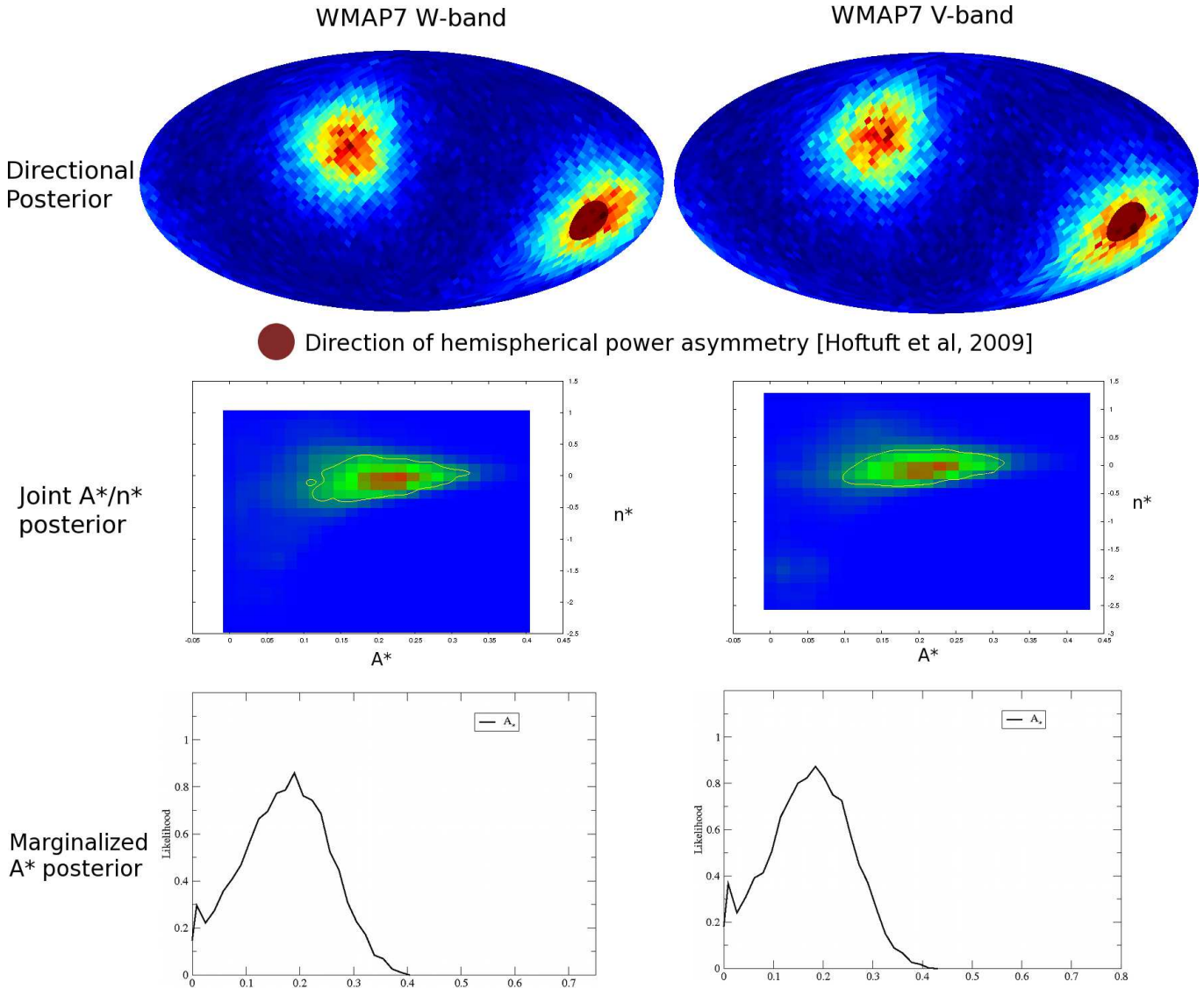, width=\linewidth,clip=}}
\caption{The main results from the seven-year W (left) and V (right)
  band WMAP data. Note that the directions are stable in all bands
  (including the Q-band), and that the anisotropy amplitude $\cA_*$ is
  nonzero at $2\sigma$ in all cases. The direction of the previously
  described hemispherical power asymmetry is marked with a red circle.}
\label{fig:main_wmap7_results}
\end{figure*}

We present the marginal posteriors for the dipole model obtained from
the seven-year WMAP temperature sky maps, as computed with the method
described in Section \ref{sec:methods}. First, in the top row of Figure
\ref{fig:main_wmap7_results} we show the preferred direction
posteriors, $P(\hat{\mathbf{n}}|\mathbf{d})$ for the W- and V band
data. The joint posterior $P(\cA_*, n*|\mathbf{d})$ is depicted in the
middle row, while the bottom row displays the marginal
posteriors,$P(\cA_*|\mathbf{d})$, and $P(n_*|\mathbf{d})$. The results
are listed in Table \ref{tab:results}.

\begin{deluxetable}{lcccc}
\tablewidth{0pt}
\tablecaption{Summary of marginal posteriors from the seven-year WMAP data \label{tab:results}}
\tablecomments{The values
  for $\cA_*$ indicate posterior mean and standard deviation. The
  ecliptic poles are located at $\pm (96^\circ, 30^\circ).$}
\tablecolumns{4}
\tablehead{ Band & $\ell$ range  & Mask & Amplitude $\cA_*$ & Direction
  $(l,b)$  }
\startdata
W1-4 & $2-400$   &    KQ85    &   $0.17 \pm 0.08$               &
$(230^{\circ}, -25^{\circ}) \pm 20^{\circ} $                       \\
V1-2 & $2-400$   &    KQ85    &   $0.18 \pm 0.08$               &
$(225^{\circ}, -25^{\circ}) \pm 20^{\circ} $                       \\
V1-2 & $2-400$   &    KQ75    &   $0.18 \pm 0.08$               &
$(225^{\circ}, -25^{\circ}) \pm 20^{\circ} $                       \\
Q1-2 & $2-400$   &    KQ85    &   $0.19 \pm 0.10$               &
$(225^{\circ}, -25^{\circ}) \pm 20^{\circ} $
\enddata
%\label{tab:results}
\end{deluxetable}

The direction of the dipole amplitude is located at about
$(l,b)=(225^\circ, -20^\circ) \pm (20^\circ,20^\circ)$ in all the
bands. This corresponds to the direction of the previously discovered
hemispherical power asymmetry by \cite{2009ApJ...699..985H} at $(l,b)
= (225^\circ,-27^\circ)$, and suggests that the signal has a
cosmological origin. The tilt parameter $n_*$ is found
consistently around $0$, implying that the dipole effects are mostly
concentrated on large scales for $\ell < 50$. The coupling strength of
the dipole contribution $\cA_*$ for various bands are listed in Table
\ref{tab:results}. Note the value of $\cA_*$ is close to zero, there are
several contaminating sources that contribute to the posterior.  One
source is the fact that there are degeneracies between $\cA_*$ and $n_*$
for low values of $\cA_*$, so the marginal posterior have noise-related
contributions close to $0$. In addition, when $\cA_*$ is close to zero,
there is a contribution from the symmetrical posterior from negative
$\cA_*$ that spill over to positive values. When keeping this in mind,
it should altogether be clear that when inspecting the posteriors that
the amplitude parameter $\cA_*$ is nonzero at about a $2.0\sigma$ significance
in the W-, V- and Q bands.

Since the canonical form of non-commutativity, given by Eq.(\ref{com})
with $\theta$ a constant in the comoving frame, is of specific
interest, we have tested this case separately.  Basically this amounts
to fixing the spectral index to the theoretical prediction $n_*=2$ and
varying only three parameters. In that case we that the data is
consistent with a vanishing anisotropic contribution and bounds $\cA
\lesssim 0.05$. This translates into a couple of orders of magnitude
looser bounds than obtained by \cite{Akofor:2008gv} from the power
spectrum alone, since the effect is at small scales which we cut out
from the analysis $\ell>400$.

\section{Conclusion}
\label{sec:conclusion}
In this paper, we have developed a numerical method for investigating
traces of non-commutative geometry derived from field-theoretical
implementations of microcausality violation in the seven-year WMAP
data. The deformation of Lorentz symmetry relevant at inflation
induces parity-violating modulations of the primordial power spectrum,
which give rise to a dipole-modulation effect in the CMB. The dipole
modulation has certain similarities to the previously detected
hemispherical power asymmetry, and not surprisingly, we reproduced the
direction of the hemispherical power asymmetry at $(l,b) =
(-225^\circ,-25^\circ) \pm (20^\circ, 20^\circ)$ when analyzing the
combined seven-year data sets. In addition, both the direction and
amplitude are stable and nonzero at a $2 \sigma$ level in the W-, V-
and Q bands. The tilt parameter $n_*$ is firmly located around zero,
indicating that the seven-year WMAP data prefers the dipole
modulations to occur on large scales $\ell < 50$. While these results
are intriguing, the significance is still too low to be considered a
clear detection.  This could be due to the fact our parameterization
of the leading order effect may not capture the underlying physics to
the full effect.  In addition, one should take into account the higher
order multipoles and vary also the cosmological parameters.

\begin{acknowledgements}
  We thank Hans Kristian Eriksen and Frode Kristian Hansen for useful
  discussions.  We acknowledge use of the
  HEALPix\footnote{http://healpix.jpl.nasa.gov} software
  \citep{gorski:2005} and analysis package for deriving the results in
  this paper. We acknowledge the use of the Legacy Archive for
  Microwave Background Data Analysis (LAMBDA). Support for LAMBDA is
  provided by the NASA Office of Space Science. The authors
  acknowledge financial support from the Research Council of Norway.
  DFM thanks Research Council of Norway FRINAT grant 197251/V30. The
  work of TK was supported by the Academy of Finland and the Yggdrasil
  grant of the Research Council of Norway.
\end{acknowledgements}

\section*{}
%\appendix

\end{document}